\documentclass{article}
\usepackage{psfig}
\usepackage{a4wide} 
\usepackage{verbatim}

\newcommand{\half}[1]{\frac{#1}{2}}

\renewcommand{\>}{\rangle} 
\newcommand{\<}{\langle} 

\newcommand{\tr}{\mbox{tr}\,}
\newcommand{\ie}{{\em i.e. }}
\newcommand{\re}{\mbox{Re}\,}
\newcommand{\im}{\mbox{Im}\,}    
\renewcommand{\to}{\rightarrow} 
 
\newcommand{\fig}[1]{Figure~\ref{#1}} 
\begin{document} 
\title{Second quantization approach to characteristic polynomials in RMT}

\author{Dimitry~M.~Gangardt 
 \\ \em Department of Physics, Technion, 32000 Haifa, Israel}  


\maketitle 





 

\begin{abstract} 
The  distribution of the characteristic polynomial $Z(U,\theta)$ of 
$N\times N$ matrices $U$ in the Circular Unitary Ensemble
is studied by the method of second quantization for one-dimensional
fermions. For infinite $N$ the Gaussian distribution of $Z(U,\theta)$ 
is established straightforwardly  by  bosonization. A general expression 
for the $n$-point correlation function of the characteristic polynomial at
different points is given by this method.  The case of finite $N$ 
is discussed.
\end{abstract} 
\noindent 
The statistical properties of the Riemann zeta function \cite{titch} 
have been extensively studied analytically \cite{montgomery,sarnak} and 
numerically \cite{odlyzko} and their analogy with 
the corresponding properties of  ensembles
of random matrices was investigated within the framework of the 
random matrix theory \cite{mehta}. 
Recently, the distribution  of values taken by the characteristic polynomials 
\begin{equation}
Z(U,\theta) = \det\left(I-Ue^{i\theta}\right)=
\prod_{j=1}^N \left(1-e^{i(\theta-\theta_j)}\right)
\label{z}
\end{equation}
of $N\times N$ unitary random matrices $U$ with eigenvalues $e^{-i\theta_j}$
belonging to the circular unitary ensemble (CUE) was investigated 
in~\cite{sk,hko}.
In particular, it was shown by explicit calculations  that in the limit
$N\to\infty$ the distribution of real and imaginary part  of
$\log Z(U,\theta)$ divided by a factor  $\sqrt{(1/2)\log N}$  
is a standard normal distribution 
in two dimensions. This was conjectured  to mimic  the similar behavior 
of the Riemann zeta function high up the critical line.
The convergence of the corresponding cumulants to the 
gaussian limit as $N\to\infty$ was also explicitly calculated and they were
conjectured to describe the corresponding properties of the Riemann
zeta function. 

In this paper I use the equivalence between  the CUE and a theory of 
fermions in one dimension to calculate the statistical properties of   
$\log Z(U,\theta)$ using the method of second quantization.
One of the most prominent features of this approach is that the case  
$N=\infty$ can be studied from the very beginning,  thus avoiding the 
tedious finite $N$ calculations and the asymptotic expansion. For 
infinite $N$ the calculations  simplify a lot, since in this case the 
fermionic theory is equivalent to a theory of free bosons, the fact known 
under the name of {\em bosonization} \cite{bos}. In what follows I calculate 
the distribution functional of the characteristic polynomial (\ref{z}) for 
infinite $N$ using this equivalence. In mathematical literature
the bosonization is known under the name of Frobenius formula for
irreducible characters of the permutation group \cite{littlewood} or  
Szeg\"o asymptotic formula for Toeplitz determinants.

Let me briefly describe the relation between the CUE and the fermions
in one dimension. In the random matrix theory one is interested mainly in 
calculating statistical averages of functions, which depend only on 
the eigenvalues $e^{-i\theta_j}$
of $U$.  Consider some  symmetric function of eigenvalues 
$f(\theta)\equiv f(\theta_1,\ldots,\theta_N)$, and its 
average, defined as
\begin{equation}
\< f\>\equiv \frac{1}{(2\pi)^N N!} \int
d^N\theta\,\, \left|D \left(e^{i\theta}\right)\right|^2 f(\theta),
\label{average}
\end{equation}
where the (Haar) measure of integration is defined with help of the 
Vandermonde determinant~\cite{littlewood,mehta}:
\begin{equation}
D(z) \equiv D(z_1,\ldots,z_N) = 
\det\left[ z_j^{N-k}\right]_{j,k=1,2,\ldots,N} = \prod_{j<k} (z_j-z_k)
\label{vandermond}
\end{equation}
Consider a quantum particle on a ring $0<\theta<2\pi$, described by  
the  wave-function of the $n$-th orbital: 
\begin{equation}
\psi_n (\theta) = \frac{1}{\sqrt{2\pi}}e^{in\theta}.
\label{wave}
\end{equation}
The Vandermonde determinant 
$D\left(e^{i\theta}\right)$ is therefore proportional to the Slater 
determinant:
\begin{equation}
\Psi(\theta)\equiv\Psi_0 (\theta_1,\ldots,\theta_N)
=\frac{1}{\sqrt{(2\pi)^N N!}}
\det\left(e^{i(N-k)\theta_j}\right)
\label{psi0}
\end{equation}
composed of particles (fermions) occupying the orbitals $n=N-k=0,\ldots,N-1$.
The proportionality factor coincides exactly with  the square root of the
normalization factor in front of the integral in (\ref{average}) and 
this average can be rewritten as a  quantum-mechanical expectation value:
\nopagebreak
\begin{equation}
\< f\> = \<\Psi_0| f\,|\Psi_0\>= \int
d^N\theta\,\, \Psi_0^*(\theta)\,f(\theta) \,\Psi_0(\theta)
\label{expval}
\end{equation}
of the operator $f$ defined as $f(\theta)$ in the coordinate 
representation. This correspondence of the RMT and one-dimensional
fermions  will be used extensively throughout the paper and in particular
the statistical average $\<\ldots\>$ and the quantum expectation value 
$\<\Psi_0|\ldots|\Psi_0\>$ will be interchanged in the course of the paper
by the virtue  of (\ref{expval}). Some application of the 
fermionic picture will be presented in what follows. 

The central object of my discussion is  the logarithm of the  
characteristic polynomial (\ref{z}):
\begin{equation}
L(\theta)=\log Z(U,\theta) = 
\sum_{j=1}^N \log \left(1-e^{i(\theta-\theta_j)}\right) .
\label{logz}
\end{equation}
Expanding the logarithm, the  equation (\ref{logz})  can be rewritten as
\begin{equation}
L(\theta)=
-\sum_{k=1}^\infty \frac{e^{ik\theta}}{k}\sum_{j=1}^N e^{-ik\theta_j}=
-\sum_{k=1}^\infty \frac{ e^{ik\theta}\rho_k}{k},
\label{logz2}
\end{equation}
where we have used the Fourier transform 
of the density operator:
\begin{equation}
\rho_k=\tr\left( U^k\right)= \sum_{j=1}^N e^{-ik\theta_j}=
\int_0^{2\pi}\! d\theta\,e^{-ik\theta}\rho (\theta),
\;\;\;\;\;\;\;
\rho (\theta)= \sum_{j=1}^N \delta (\theta-\theta_j)=
\frac{1}{2\pi}\sum_{k=-\infty}^{+\infty} e^{ik\theta}\rho_k. 
\label{density}
\end{equation} 
In order to calculate the statistical properties of $L(\theta)$ using 
the correspondence (\ref{expval}) between statistical average and quantum 
expectation value it is convenient to employ the method of 
second quantization. We introduce creation and annihilation operators 
$c^\dagger_n$ and $c_n$ for a fermion on the $n$-th orbital with usual
anti-commutation relations:
\begin{equation}
\{c_n,c^\dagger_m\}\equiv  c_n c^\dagger_m+c^\dagger_m c_n=\delta_{n,m},\;\;\;\;\;
\{c_n,c_m\} =
\{c^\dagger_n,c^\dagger_m\} = 0
\label{anticom}
\end{equation}
The quantum state $|\Psi_0\>$ is then defined by the action of the creation 
operators on the vacuum:
\begin{equation}
|\Psi_0\> = \prod_{n=0}^{N-1} c^\dagger_n|\mbox{Vac}\> .
\label{creatpsi0}
\end{equation}
%
%
The second-quantized form of the  density operator 
$\rho_k$ is given by the standard rules~\cite{negorl}:
\begin{equation}
\rho_k = \rho^\dagger_{-k}= \sum_{n=-\infty}^{+\infty} c^\dagger_{n-k} c_n
\label{rhok2}
\end{equation}
for $k\neq 0$, while for $k=0$ the definition (\ref{density}) gives 
$\rho_0 = N$ --- the total number of particles. 
When acting on the state $|\Psi_0\>$ the density operator $\rho_k$ 
creates a linear  combination of states in which one particle on the $n$-th
orbital  
is moved $k$ orbitals down, 
provided the orbital $n-k$  is empty. This simple 
observation  yields the important result (for $k\neq 0$):
\begin{equation}
\<\rho_k \rho^\dagger_p\>=\delta_{k,p} C_2 (k)=
\delta_{k,p}\<\Psi_0|\rho_k \rho^\dagger_k|\Psi_0\> = \delta_{k,p}\times
\left\{\begin{array}{cc}
|k|, & |k|\le N \\
N, & |k| >N\end{array}
\right.
\label{denscorr}
\end{equation}
otherwise obtained by using the Wick theorem.
The correlation function $C_2 (k)=\<\rho_k \rho^\dagger_k\>$ with
$C_2 (0)=N^2$,   
is nothing but a Fourier transform of DOS-DOS
correlation function \cite{mehta} of CUE :
\begin{equation}
C_2 (\theta ; N)=\<\rho(\theta) \rho^\dagger (0)\>= 
\frac{1}{2\pi}\sum_{k=-\infty}^{+\infty} e^{ik\theta}C_2(k ; N)= 
\frac{N}{2\pi}\delta(\theta)+\left(\frac{N}{2\pi}\right)^2-
\left(\frac{\sin \half{N\theta}}{2\pi\sin\half{\theta}}\right)^2 .
\label{R2}
\end{equation}
The symmetry of the  correlation function  $C_2 (k ; N)$ with 
respect to $k\to -k$ follows from the fact that for finite $N$ the density 
operators $\rho_k$ and $\rho^\dagger_k$ commute. For infinite~$N$, 
as we shall see, this is not true. 

I now proceed to calculate the correlation functions of the logarithm
of the characteristic polynomial. In the work \cite{sk} the correlation 
functions
of $L(\theta)$ were calculated at the same point. 
Here I generalize this result, 
to the case of the  two-point correlation function, and calculate
\begin{eqnarray}
 K_2 (\theta_1-\theta_2 ; N) &=& \< L (\theta_1) L^\dagger (\theta_2)\> 
\label{K2}\\
 P_2 (\theta_1-\theta_2; N) &=& \< L (\theta_1) L (\theta_2)\>. 
\label{P2}
\end{eqnarray} 
where in the coordinate representation $ L^\dagger (\theta)= L^* (\theta)$.

Due to the translational symmetry
all the correlation functions depend on the 
difference $\theta=\theta_1-\theta_2$ only. 
From (\ref{K2}) and (\ref{P2}) the correlation functions of real and imaginary
part of $L(\theta)$ can be easily obtained.
I notice that $P_2 (\theta)$ vanishes due to the fact that the 
Kronecker delta in (\ref{denscorr}) is never satisfied. 
Moreover, it follows that
\begin{eqnarray}  
Q_2 (\theta_1-\theta_2 ; N) &\equiv& \< \re L(\theta_1)\, \re L(\theta_2)\> = 
\half{\re K_2 (\theta_1-\theta_2 ; N)} \nonumber \\
R_2 (\theta_1-\theta_2 ; N) &\equiv & \< \im L(\theta_1)\, \im L(\theta_2)\> = 
\half{\re K_2 (\theta_1-\theta_2 ; N)}
\label{Q2}
\end{eqnarray}
and in addition there exists a cross-function given by
\begin{equation}  
X_2 (\theta_1-\theta_2 ; N) \equiv \< \im L(\theta_1) \, \re L(\theta_2)\> = 
- \< \re L(\theta_1)\, \im L(\theta_2)\>=
\half{\im K_2 (\theta_1-\theta_2 ; N)}
\label{X2}
\end{equation} 
It can be checked that
$\< L(\theta)\>=\<L^\dagger (\theta)\>=0$ so there is no difference 
between connected and disconnected correlation functions. 

The calculation of $K_2 (\theta ; N)$ using the definition (\ref{logz2}) 
is straightforward:
\begin{equation}
 K_2 (\theta_1-\theta_2 ; N) = 
\sum_{k,p=1}^{\infty} \frac{e^{ik\theta}}{kp}\<\rho_k \rho^\dagger_p\>=
\sum_{k=1}^\infty \frac{e^{ik\theta}}{k^2} C_2 (k ; N) = 
\sum_{k=1}^N \frac{e^{ik\theta}}{k}+
N \sum_{k=N+1}^\infty\frac{e^{ik\theta}}{k^2}
\label{K2calc}
\end{equation}
For correlations at the same point, $\theta=0$,  the result is real and 
its half coincides precisely  with the expression (43) in~\cite{sk}:
\begin{equation}
\half{K_2 (0;N)} = Q_2 (0; N)=R_2 (0; N) = \half{1}\sum_{k=1}^N \frac{1}{k}+
\half{N} \sum_{k=N+1}^\infty\frac{1}{k^2}
\label{K20}
\end{equation}
which behaves as $(1/2)\log N$ for large $N$.
It is worth noticing
that for $\theta\neq 0$ the function $K_2 (\theta)$ is complex, therefore
there exists correlation between real and imaginary parts of $L(\theta)$
at different points, a fact which is missed when  the correlation
function is calculated at the same point.

Now the calculations for the whole distribution of 
$L(\theta)$ and $L^\dagger(\theta)$, equivalent to the distribution of
$\re L(\theta)$ and $\im L(\theta)$ is presented. 
In order to be able to calculate general $n$-point correlation functions, 
one has to consider the following generating functional:
\begin{equation}
\Xi[s,s^*] = \left\< \exp \left\{\int_0^{2\pi} \frac{d \theta}{2\pi}\left[
s^*(\theta)L(\theta)+s(\theta)L^\dagger(\theta)\right]\right\}\right\>
\label{Xi}
\end{equation}
where $s(\theta)$ and $s^* (\theta)$ are the source terms.
Any  $n$-point correlation function can be represented as a functional 
derivative of $\Xi[s,s^*]$:
\begin{equation}
\< L(\theta_1) L(\theta_2)\ldots L^* (\theta_m)  
L^* (\theta_{m+1})\ldots\> =\left. 
\frac{\delta}{\delta s^*_1}\frac{\delta}{\delta s^*_2}
\ldots\frac{\delta}{\delta s_m}\frac{\delta}{\delta s_{m+1}}
\ldots \Xi [ s,s^*]\right|_{s,s^*=0}
\label{funcder}
\end{equation}
where $s_m$, $s^*_l$ stand for $s (\theta_m)$ and $s^* (\theta_l)$
respectively.
Using the Fourier transform of these source terms
\begin{equation}
s_k = \int_0^{2\pi} d\theta\, e^{-ik\theta} s (\theta)\;,\;\;\;\;\;\;
s^*_k =(s_k)^*
\label{fours}
\end{equation}
and the expression (\ref{logz2}) the generating functional (\ref{Xi}) can be
rewritten as
\begin{equation}
\Xi[s,s^*] = \left\< \exp \left\{-\sum_{k=1}^\infty \frac{1}{k} 
\left(s^*_k \rho_k+s_k\rho^\dagger_k\right)\right\}\right\> .
\label{Xik}
\end{equation} 
It will be calculated in the limit $N=\infty$. 
It is convenient to redefine the numbering
of one-particle orbitals so the upper  occupied level  in $|\Psi_0\>$
corresponds now to $n=0$ and all the states with $n \le 0$ are occupied.
This state is the infinitely deep Fermi sea --- the ground state of the 
fermionic system, if  one-particle energy $E_n$ is an increasing function
of the level index $n$.
In addition, this state is now annihilated by the action of $\rho_k$ for 
$k>0$: it is impossible to promote a fermion from the state $n$ to an 
empty lower state $n-k$.  It is well known from the theory of 
one-dimensional correlated electrons \cite{haldane} that the operators 
$\rho_k$ and $\rho^\dagger_k$ acquire non-zero commutation relations in 
the presence of infinite filled Fermi sea (Schwinger terms). To show it
let us  begin with
\begin{equation}
[\rho_k, \rho^\dagger_p]=
\sum_{n,m=-\infty}^{+\infty} [c^\dagger_{n-k}c_n,c^\dagger_{m+p}c_m] = 
\sum_{n=-\infty}^{+\infty}\left(c^\dagger_{n-k}c_{n-p}-
c^\dagger_{n-k+p}c_n\right)\;.
\label{comm}
\end{equation}
The result is nonzero, since the operators are not well-behaved. It is 
necessary to extract the singular part by introducing the {\em normal ordered}
operators in the state $|\Psi_0\>$  by extracting the expectation value 
in this state:
\begin{equation}
:c^\dagger_m c_n: = c^\dagger_m c_n-\<c^\dagger_m c_n\> = 
c^\dagger_m c_n-\delta_{m,n} n^0_m\;,
\label{normal}
\end{equation}
where $n^0_m \equiv \<c^\dagger_m c_m\>$ is the Fermi-Dirac distribution
of occupation numbers in the ground state. Rewriting 
$c^\dagger_m c_n = :c^\dagger_m c_n:+\delta_{m,n} n^0_m$ and substituting it 
into (\ref{comm}) the commutation relation of density operators becomes
\begin{equation}
[\rho_k, \rho^\dagger_p]=
\sum_{n=-\infty}^{+\infty}\left(:c^\dagger_{n-k}c_{n-p}:-
:c^\dagger_{n-k+p}c_n:\right)  + 
\delta_{k,p}\sum_{n=-\infty}^{+\infty}
\left(n^0_{n-k}-n^0_{n}\right) = k \, \delta_{k,p} \;.
\label{commcalc}
\end{equation}
In this expression the normal ordered operators were canceled, since they 
are not singular. The last sum equals to the number of orbitals from
$1$ to $k$.

When calculating the matrix elements as in (\ref{Xik})
the order of $\rho_k$ and $\rho^\dagger_k$ is important when these operators  
do not commute.
One requires that the matrix element should coincide with the statistical
average in the coordinate representation (\ref{average}). Suppose that an 
operator $\rho_k$ for $k>0$ happens to be next to the left of the state
$|\Psi_0\>$. The result would be zero, since this state is annihilated 
by $\rho_k$ for $k>0$, which is not true for the statistical average. In
order to obtain the correct expression all operators $\rho_k$ for $k>0$
must be placed to the left of $\rho^\dagger_k$ for $k>0$, in the 
so-called anti-normal order. The operators in the
generating  functional (\ref{Xik}) must be  therefore rearranged as
\begin{equation}
\Xi[s,s^*] =
\left\< \exp\left(-\sum_{k=1}^\infty \frac{s^*_k \rho_k}{k}\right) 
 \exp\left(-\sum_{k=1}^\infty \frac{s_k \rho^\dagger_k}{ k}\right)  \right\>
\label{Xiorder}
\end{equation}    
Using the fact that the commutator of $\rho_k$
and $\rho^\dagger_k$ is a $c$-number and $\rho_k$ annihilates the ground state
for $k>0$, the generating functional  is given by the 
famous Baker-Campbell-Hausdorff formula:
\begin{eqnarray}
\Xi[s,s^*] &=& 
\left\< \exp\left(-\sum_{k=1}^\infty \frac{s^*_k \rho_k}{k}\right) 
 \exp\left(-\sum_{k=1}^\infty \frac{s_k \rho^\dagger_k}{ k}\right)\right\> = 
\exp\left(\sum_{k=1}^\infty\frac{s^*_k s_k}{k^2}
[\rho_k,\rho^\dagger_k]\right)
\nonumber\\
&=&
\exp\left(\sum_{k=1}^\infty\frac{s^*_k s_k}{k}\right)
\label{Xikcalc}
\end{eqnarray} 
which is obviously gaussian. In particular,  taking the corresponding
derivatives of  $\Xi[s,s^*]$ with respect to   $s_k/k$ and $s^*_k/k$
in analogy to (\ref{funcder})  the correlation 
function of powers of of $\rho_k$ or  traces of  $U^k$ can be calculated:
\begin{equation}
\left\<\prod_{k}{\rho^\dagger_k}^{m_k} 
\prod_p \rho_{p}^{m'_p}\right\> = 
\left\<\prod_{k}\left(\tr {U^\dagger}^k\right)^{m_k} 
\prod_p \left(\tr U^p\right)^{m'_p}\right\>=\prod_k 
\delta_{m_k,m'_k} k^{m_k} m_k!
\label{diaconis}
\end{equation}
in accordance with the results of \cite{ds}. This result and  the gaussian  
characteristic functional (\ref{Xikcalc}) are consequences of the
anomalous commutation relations  (\ref{commcalc}) that are fulfilled  
by the operator $\rho_k$ 
and its hermitean conjugate and the fact that $\rho_k$ 
annihilates the ground state for $N=\infty$. It is worth mentioning here that
 the  correlation function (\ref{diaconis}) and its generating functional 
(\ref{Xikcalc}) are in fact the restatements in terms of the field theory
of the so called functional central limit theorem for $\log Z$ discussed 
in \cite{hko}. There it was proven using the Frobenius formula for the 
irreducible characters of the permutation group, which constitutes the 
mathematical basis of the bosonization \cite{bos}.

Returning to the function $s(\theta)$ of the angle, the bilinear form in the
exponent of (\ref{Xikcalc}) can be rewritten as a double integral
\begin{equation}
\log \Xi[s,s^*] = \int_0^{2\pi}\int_0^{2\pi}\frac{d\theta_1}{2\pi} 
\frac{d\theta_2}{2\pi}\,
s^* (\theta_1)\, K_2 (\theta_1-\theta_2;\infty)\,s (\theta_2) \, ,
\label{logXi}
\end{equation}
where the ``propagator'' is given by (\ref{K2calc}) for infinite $N$, \ie
\begin{equation}
K_2 (\theta;\infty) = \sum_{k=1}^\infty \frac{e^{ik\theta}}{k}e^{-2k\eta} = 
-\log \left(1-e^{i\theta} e^{-2\eta}\right)
\label{K2inf}
\end{equation}
and the infinitesimal imaginary part $\eta$ of the angle $\theta$
was  included to ensure convergence at $\theta = 0$. It was not necessary 
for finite $N$, since in (\ref{K2calc}) the first sum is finite and the 
second one is absolutely convergent. It is well known \cite{bos,haldane} that
an ultraviolet cut-off $1/\eta$ should be introduced when dealing 
with an infinite Fermi sea.  Using this cut-off, the correlation function 
(\ref{K2inf})  at the same  point ($\theta=0$) is described  by
\begin{equation}
K_2 (0;\infty)  = 
-\log \left(1-e^{-2\eta}\right)\approx\log\frac{1}{2\eta} , 
\label{K20inf}
\end{equation}  
which should be compared with the leading behavior of
$K_2 (0;N)\sim \log N$. In fact, for large but finite $N$
one can make the {\em scaling ansatz} for the correlation function 
$K_2 (0;N,\eta)$:
\begin{equation}
K_2 (0;N,\eta) = \log\frac{1}{2\eta}
\times
f\left({\log N}/{\log\frac{1}{2\eta}}\right)
\label{scaling}
\end{equation}
with $f(x)\rightarrow 1$ as $x\rightarrow\infty$ and $f(x)\approx x$ for 
$x\rightarrow 0$. The scaling form (\ref{scaling}) can be justified by 
calculating the leading behavior of  (\ref{K2calc}) for $N \rightarrow\infty$
and $\theta=2i\eta\rightarrow 0$. Using the scaling (\ref{scaling}) the 
correspondence
between $N$ and $1/2\eta$ is established  for the leading behavior
(in $\log N$) of the second  moments  of  $\re L(\theta)$ and 
$\im L(\theta)$ at the same point. Whether the bosonization, which 
applies for the case $N=\infty$ only, can provide results for other correlation
functions for finite $N$ through some scaling relations like (\ref{scaling})
is an open question.

\begin{figure} 
\centerline{\psfig{figure=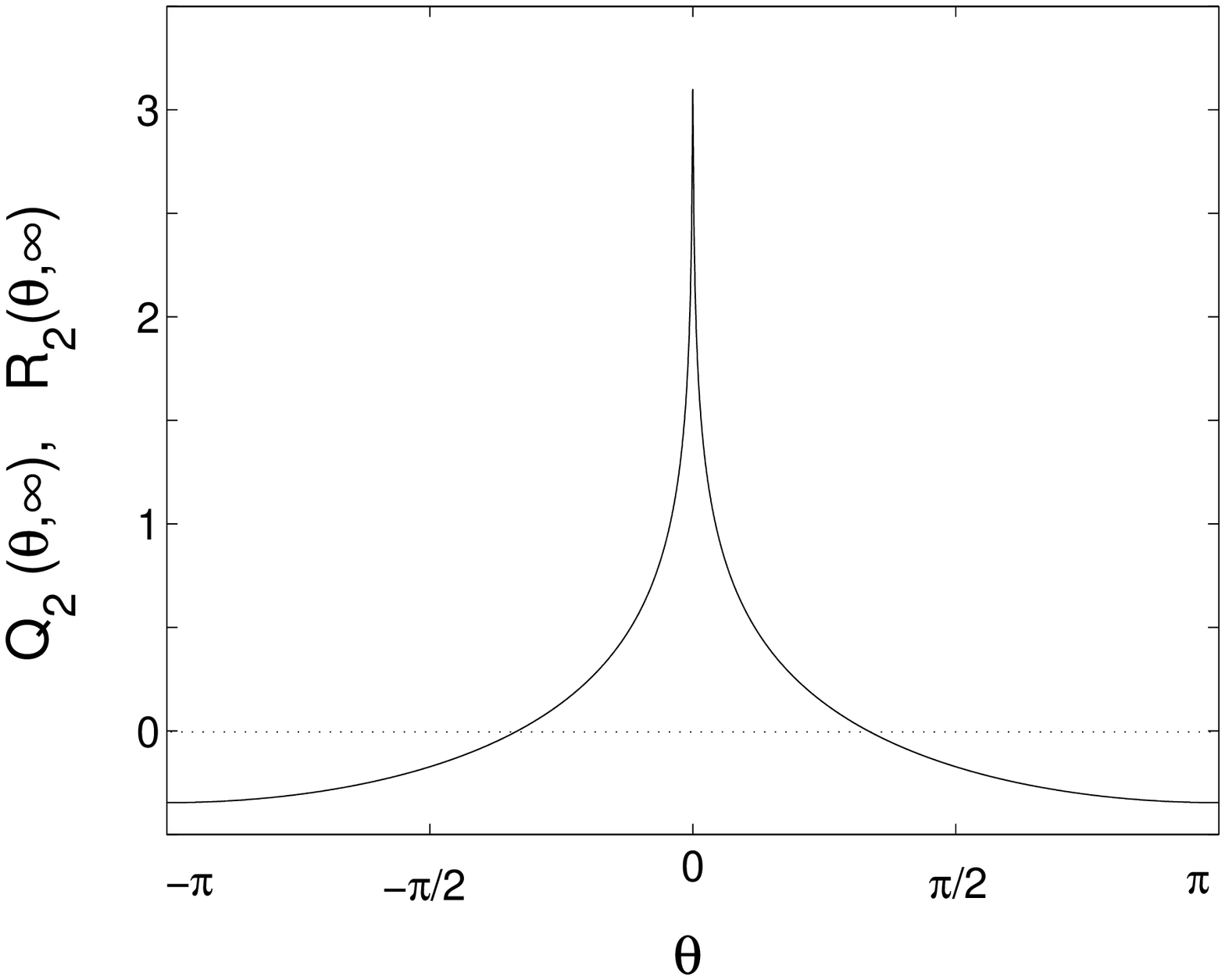,height=5cm} 
\psfig{figure=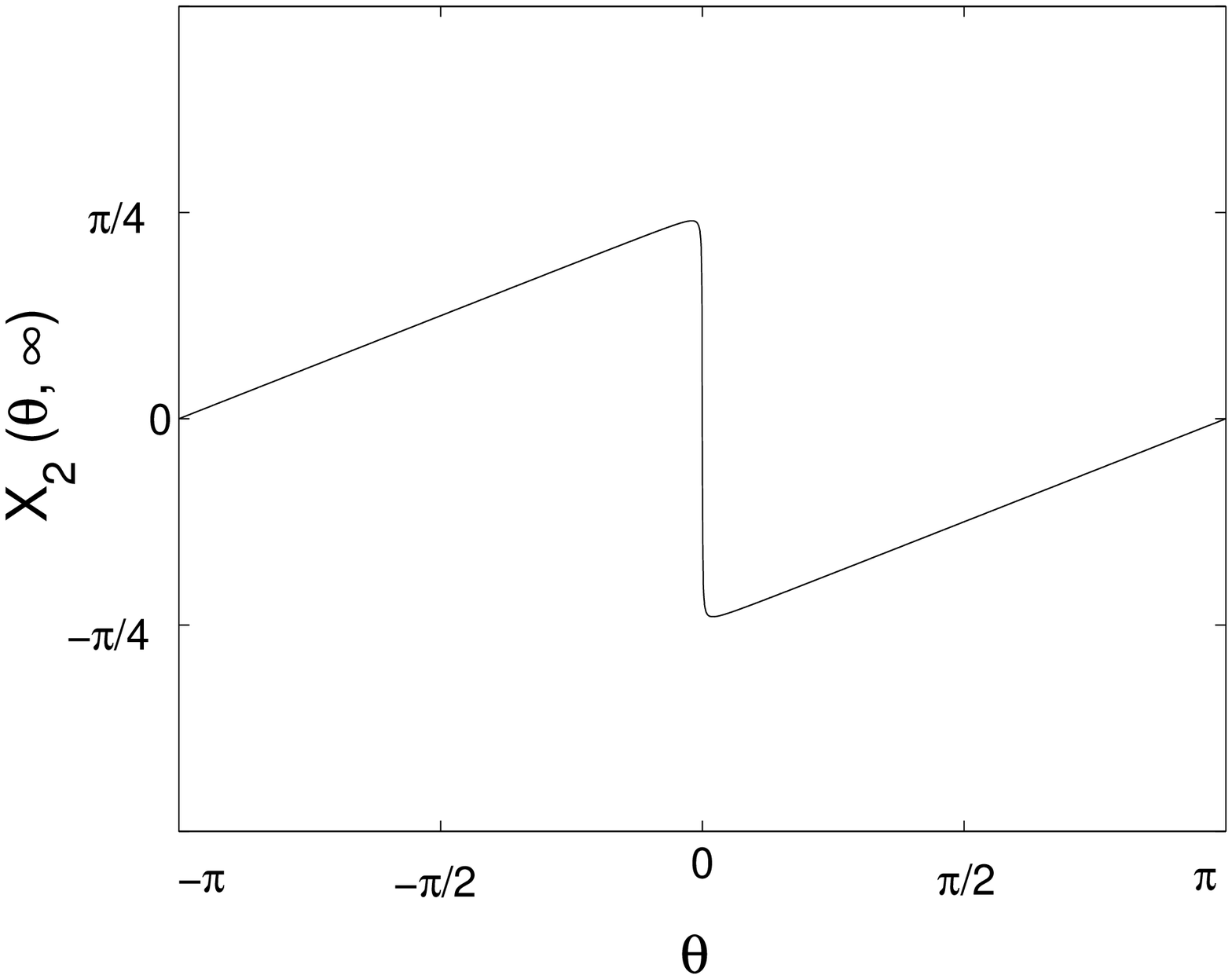,height=5cm}} 
\caption{The correlation functions $Q_2 (\theta ; \infty)$, 
$R_2 (\theta ; \infty)$ and $X_2(\theta ; \infty)$ for $\eta= 0.001$.} 
\vspace{0.5cm}
\label{q2x2} 
\end{figure}

Finally I give explicit expressions for correlation functions 
$Q_2$, $R_2$ and $X_2$ for $\theta\neq 0$ calculated using 
(\ref{Q2},\ref{X2}). 
Separating the real and imaginary parts in (\ref{K2inf})  they read:
\begin{eqnarray}
&&Q_2 (\theta ; \infty )=R_2 (\theta ; \infty) = 
\half{1}\log\left(\frac{1}{\left|2\sin\half{\theta}\right|}\right) 
\label{Q2calc}\\
&&X_2 (\theta ; \infty ) = \frac{\theta-\pi}{4}\;\mbox{mod}\, \half{\pi}
\label{X2calc}
\end{eqnarray}
where the $\mbox{mod}\,\pi/2$ means that the values taken by 
$X_2 (\theta ;\infty )$ lie
in the interval $[-\pi/4,\pi/4]$ so it is a periodic function of the angle.
The functions (\ref{Q2calc},\ref{X2calc}) are shown in \fig{q2x2}. It would 
be interesting to compare these correlation functions with the corresponding 
functions for real and imaginary parts of the Riemann zeta function high up 
the critical line. The function $R_2 (\theta; \infty)$
is related to the variance of number of eigenvalues in the arc of length 
$\theta$ which was  studied in \cite{rains}.

In this paper I have shown how the one to one correspondence between the 
probability measure and Slater determinants of fermions in one  dimension
can be used in order to calculate different statistical properties of 
Circular Unitary Ensemble of random matrices. In particular, the correlation
functions of the density of states, which corresponds to the density of the 
fermions are obtained with help of the Wick theorem as a diagrammatic 
expansion. This method was applied to the calculation of correlation functions
of characteristic polynomials.  The case of $N=\infty$ was treated by 
the method of bosonization  and  the generating functional of correlation 
functions was calculated exactly yielding an alternative derivation
of some of the results of \cite{sk,hko,rains,ds}. 
The corresponding distribution was shown to be gaussian corresponding
to the functional central limit theorem of \cite{hko}.
In the future it would be interesting to investigate by the present method
the case of finite $N$ in order to obtain the correlation
functions of the characteristic polynomials at the same point. 

In conclusion I would like to remark that the finite $N$ calculations 
presented in this paper and  in \cite{sk,hko,rains,ds} are related to  
the results obtained for the (static) correlation functions of  strongly 
correlated particles in one dimension. Indeed,  it is known  that the 
different circular ensembles of random matrices correspond to the 
Calogero-Sutherland model \cite{cs}, 
which describes a system of $N$ interacting particles. This model with  
interactions falling off as the inverse square of the  distance between the 
particles and proportional to the coupling constant $\lambda$, was 
shown in \cite{ha} to be equivalent to the system of 
effective free particles with  fractional quantum  statistics.
The values of the  coupling constant $\lambda=1/2$, $1$ and $2$ correspond 
to  the orthogonal, the unitary and the symplectic ensemble respectively. 
In the present work only the  $\lambda=1$  Calogero-Sutherland model
equivalent to  the free fermions was considered  within the framework of 
the second quantization. The generalization of the present approach to other 
ensembles, corresponding to more exotic fractional statistics would be an
interesting issue.

I would like to thank 
BRIMS Hewlett-Packard Labs in Bristol for hospitality.
I am grateful to J.P.~Keating   for fruitful, stimulating 
discussions.  I am also grateful 
to S. Fishman for his comments and generous  help during the preparation
of this manuscript.
This research was supported in  part by  
the U.S.--Israel Binational Science Foundation (BSF), by the Minerva  
Center for Non-linear Physics of Complex Systems, by the Israel  
Science Foundation, by the Niedersachsen Ministry of Science  
(Germany) and by the Fund for Promotion of Research at the  
Technion.

\end{document}